\documentclass[floatfix,amsmath,amsfonts,twocolumn]{revtex4}  
\usepackage{enumitem}
\usepackage{amsmath}
\usepackage{graphicx}
\usepackage{epsfig}
\usepackage{bm}
\usepackage{bbold}
\usepackage{amssymb}
\usepackage{color}

\newcommand{\px}{{\partial_x}}

\newcommand{\pt}{{\partial_t}}

\newcommand{\RE}{{\operatorname{Re}}}

\newcommand{\z}{\zeta}
 
\newcommand{\rev}[1]{\textcolor{black}{#1}}

\begin{document}

\title{Bright and dark  solitons in the systems with strong light-matter coupling: exact solutions  and  numerical simulations 
%Light-matter bright and dark solitons: exact solutions and stability
}

\author{A. V. Yulin and  D. A. Zezyulin\footnote{email: d.zezyulin@gmail.com}}

\affiliation{ITMO University, St.~Petersburg 197101, Russia}
	%\\
	%$^{2}$  Moscow Institute of Electronic Engineering, Zelenograd, Moscow, 124498, Russia

\date{\today}

\begin{abstract}
    We  theoretically study bright and dark solitons in an experimentally relevant hybrid system   characterized by strong light-matter coupling. We find that the corresponding two-component model  supports a variety of coexisting moving    solitons including   bright solitons on zero and nonzero background, dark-gray and gray-gray dark solitons.   The solutions are found in the analytical form by reducing the two-component problem to a single stationary equation with cubic-quintic nonlinearity.   All  found solutions coexist  under  the same set of the model parameters, but, in a properly defined linear limit,   approach different branches of the polariton dispersion relation for linear waves. Bright solitons with zero background feature an oscillatory-instability threshold which can be associated with a resonance between the    edges of the continuous spectrum branches. `Half-topological' dark-gray and nontopological gray-gray solitons are stable in wide parametric ranges below the modulational instability threshold, while bright solitons on the constant-amplitude pedestal are unstable.  

%\emph{Keywords:}  nonlinear Schr\"odinger, solitary wave, localized, absorption, dissipation, soliton family

\end{abstract}

\maketitle

\section{Introduction}

%It is well known that optical waves propagating in nonlinear system can show very rich dynamics and has been studied for many years. One of the intriguing and practically important phenomenon occurring in the systems is the formation of solitary waves and  patterns. Roughly speaking the solitons appear due to perfect counter balancing of the spreading of the pulse by the effects leading to the compression of the pulse.  Intense investigation of solitary waves was triggered by the discovery that localized solutions of integrable equation restore their shapes completely after collisions with other waves envelopes. Perturbations destroying intergability of the system allow the solitons to change their energy but in many cases the solitons show themselves to be very robust against weak perturbations. This makes it convenient to describe the evolution of the system in terms of the solitons with parameters (such as energy or velocity) varying in time. 

Optical solitons, as localized waves propagating in nonlinear fibers, were predicted \cite{Hasegawa} and experimentally observed  \cite{Mollenauer} more than forty years ago. Since then optical solitary waves have  been discovered and thoroughly  studied in many optical systems. Apart from the fundamental importance, optical solitons can also be of interest from practical point of view, in particular,     for information transmission and   supercontinuum generation  \cite{Haus, Dudley, Skryabin}. 
%It is worth noting here that initially the term "soliton" was introduced to  denote the localized solution of integrable equations but then was extended to all nonlinear localized structures including dissipative ones.  
To describe formation of solitons, %the evolution of light 
it is necessary to combine Maxwell equations with the equations accounting for the response of the  material   to the propagating 
%describing the properties of the material affected by the 
electromagnetic field. This description is, however, in many cases so complicated that even numerical modelling of the dynamics of the light becomes impossible. Fortunately, the presence of small parameters often allows to simplify the problem. For example, slow varying amplitude approximation has proven to be a very powerful and precise model which allows to describe  a large variety of optical phenomena using the nonlinear Schr\"odinger (NLS) equation and a family of its generalizations \cite{AA,Kivshar}.  In its basic form, the NLS equation is fully integrable and its soliton  solutions are available in the analytical form.  The knowledge of exact soliton solution for  the NLS equation and other prototypical nonlinear   models has  two-fold importance.   First, it helps to understand the properties of the localized waves in the considered system and, in particular, facilitates the stability study. Second, the analytical  solutions can be used as a starting point for the development of a perturbation theory for more complex and general systems which take into account   the originally neglected effects and do not always admit analytical solutions. 
%It is also important to know whether the solitary pulses are stable or not. Let us remark that, of course, from physical point of view, the stable solitons are of most interest because they can be observed in experiments. However mathematically the unstable solitons can also be of useful. 

%One of the basic model describing evolution of pulses in nonlinear waevguiding systems is the  nonlinear Schr\"odinger equation. A soliton solution of this equation is well known as Schrodinger Soliton.  As it is mentioned above the soliton solution is often very robust against small perturbations and the evolution of the system described by a Generalized Nonlinear Schrodinger Equation ( NLS with some additional terms ) can often be understood in terms of dynamics of interacting Schrodinger solitons.       

%So the knowledge of exact soliton solution of the basic models may have two-fold importance.   First, it helps to understand the properties of the localized waves in the considered system. Secondly, these soliton solutions can be used as a starting point for the development of a perturbation theory for more complex systems. It is also important to know whether the solitary pulses are stable or not. Let us remark that, of course, from physical point of view, the stable solitons are of most interest because they can be observed in experiments. However mathematically the unstable solitons can also be of useful. 

In this paper, we address bright and dark %conservative 
solitons propagating in optical waveguides with strong light-matter coupling which is a typical attribute of exciton-polariton systems. The  system of such a kind consist of a dielectric waveguide with built-in quantum wells supporting excitons. If the losses are small, then, at the frequencies close to the exciton resonance, the photons and the excitons interact strongly so that at the crossing point the dispersions of the photons and the excitons hybridize and %, their dispersions 
get split forming the lower and the upper polariton branches. 

Without resonant material excitations, the nonlinear effects usually come to play at so high pulse energies that it reduces   its practical applicability, especially in optical on-chip devices. The effective (material and waveguide) dispersion is also relatively low in these systems. For example, typical energy of $100$\,fs optical solitons in highly nonlinear optical waveguides is  of the order of $100$\,pJ, and the soliton formation occurs at propagation distances of several centimeters. The advantage of the systems with strong light-matter coupling is that because of the material component of the eigenmodes the nonlinear effects are orders of magnitude stronger than in the systems with weak coupling. Another important fact is that the typical dispersion caused by the linear photon-exciton interaction is much stronger compared to the dispersion of pure photons. This allows   to observe soliton formation at the propagation distances of order of hundreds microns and the energy of order of hundreds of fJ in $100$\,fs pulses. This makes the systems with strong light-matter interaction to be very promising for studying different nonlinear effects and explains why these  systems  have been attracting so much of attention over the  recent years \cite{CarusottoCiuti}.   

The simplest model describing the dynamics of pulses propagating in the waveguides with strong light-matter interaction neglects the dispersion and nonlinearity of pure photons, and  hence the photonic component is described by a linear equation. This equation is coupled to the   equation for   excitons where the resonant frequency of the excitons is a function of their density. In the simplest case considered herein, the shift of the exciton frequency is proportional to their density. In the context of meanfield approximation for polariton systems, the corresponding model was introduced in \cite{Carusotto} and has been widely used. The adaptation of the model for the case when the frequency of excitation is much higher than the cut-off frequency of the waveguide is done in \cite{NatCom15}. The latter model can be brought to the following form: 
%\begin{equation}
%	\label{eq:main}
%\begin{array}
%i(\pz A + \pt A) = -\kappa \psi,\\[2mm]
%i\pt \psi = -\kappa A + g |\psi|^2\psi.
%\end{array}
%\end{equation}
\begin{equation}
	\label{eq:main}
i(\pt A + \px A) = -\kappa \psi, \quad i\pt \psi = -\kappa A + g |\psi|^2\psi.
\end{equation}
Here $t$ is the time normalized on some characteristic frequency $\Omega_{0}$, $x$ is the coordinate normalized on the $\Omega_0/v_g$, $v_g$ is the group velocity of the pure photon mode at the resonant frequency of the material excitations $\Omega_m$.  In the equations (\ref{eq:main}) the coefficient $\kappa$ accounts for the light-matter coupling strength and  without loss of generality can be set to $1$. This means that the normalization frequency $\Omega$ is chosen to be equal to the light-matter coupling in the system.

The function  $A(x,t)$ in (\ref{eq:main}) is the slow varying amplitude of the photon field and $\psi(x,t)$ is the order parameter describing the material excitations, for example $\psi$ can be the order parameter of the coherent excitons in a semiconductor microcavity. To achieve analytical solutions, we disregard the dispersion of the pure guided photons assuming this is much smaller compared to the dispersion appearing due to the light-matter coupling (as it is typical   for experimental conditions \cite{NatCom15}). The effective  mass of the coherent excitons is supposed to be much greater than the effective mass of guided photons (for semiconductor microcavities the typical values $<10^{-4}$) and thus the frequency of linear material excitations does not depend on their wave vector. Let us remark that all frequencies used below in this paper are actually the detunings of the frequency from the linear frequency of the material excitations $\Omega_m$. The wavevectors, in their turn, are the detunings of the wavevectors from the wavevector of the pure photonic mode of the frequency $\Omega_m$. 

Normally in the systems with strong light-matter coupling the dominating nonlinearity originates from the resonant frequency dependency of the material excitations on their density. In our model this frequency shift is taken to be proportional to the density of the material excitations and thus is equal to $g|\psi|^2$. %Finally, we would like to acknowledge that 
Aiming to obtain analytical solutions, we consider a conservative problem. The conservative limit is a good approximation for polariton waves propagating over the distances sufficient to observe the formation of the solitons (hundreds of microns). The comprehensive studies of the effect of losses is definitely of interest but is out of the scope of the present paper and will be done elsewhere.

\begin{figure}%[t]
	\begin{center}
		\includegraphics[width=0.8\columnwidth]{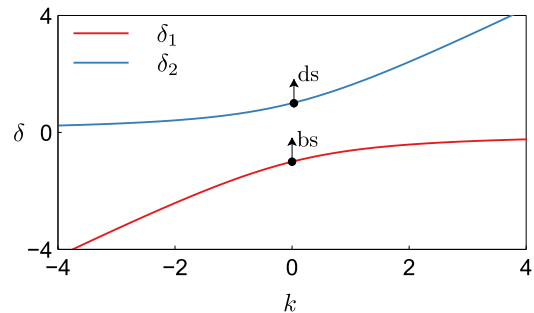}%
	\end{center}
	\caption{Dispersion relation for linear waves $\delta_{1,2}$ given by (\ref{eq:dr}). Arrows labelled  `bs' and `ds' correspond to the frequencies of, respectively, bright and dark solitons detaching from the polariton branches. Direction of arrows corresponds to the increase of the soliton amplitudes. }
	\label{fig:dr}
\end{figure}

In the linear limit $g=0$, for plane-wave solutions $\propto e^{-i\delta t + ikx}$, where $\delta$ is the frequency and $k$ is the wavenumber, we obtain two polariton branches (upper, with subscript 1, and lower, with subscript 2) of the dispersion relation (plotted in Fig.~\ref{fig:dr}):
\begin{equation}
\label{eq:dr}
    \delta_{1,2}(k) = (k\pm\sqrt{k^2+4\kappa^2})/2.
\end{equation}

System (\ref{eq:main})  can be reduced to  the  NLS equation written    for the amplitude of the polariton mode belonging either to the lower or the upper branch provided the peak power of the pulse causes the exciton frequency shift much less compared to width of the gap between the upper and the lower polariton branches. However the soliton of higher intensities have to be considered taking into account the dispersion of the nonlinearity and the excitations belonging to both branches. That is why the analysis of the full two-component model is important. We also note that a generalized version of model (\ref{eq:main}) has been used to describe the formation of dark solitons in polariton fluids \cite{Walker}.  Another pertinent remark is that system (\ref{eq:main}) is  mathematically similar to the coupled-mode equations describing   gap solitons in optical fibers with grating \cite{deSterke}. The importance difference is that in (\ref{eq:main}) only the equation for $\psi$ field (order parameter function describing coherent polaritons) is nonlinear whereas in the case of gap solitons both fields are nonlinear.

%The starting point for our analysis is system (3)-(4) in \cite{NatCom15}.  After some rescalings, we can reduce it   to the form
%\begin{subequations}
%	\label{eq:main}
%\begin{eqnarray}
%i(\pz A + \pt A) = -\kappa \psi,\\
%i\pt \psi = -\kappa A + g |\psi|^2\psi,
%\end{eqnarray}
%\end{subequations}
%where without loss of generality we assume $\kappa\geq 0$.

It has been found in \cite{NatCom15} that system (\ref{eq:main}) admits an analytical bright soliton solution whose existence has been confirmed experimentally. In this paper, we perform a comprehensive study of different kinds of bright and dark solitons existing in the system and look into stability of the found analytical solitons. 
We find that apart from the bright solitons, the system admits `semitopological' dark-gray solitons,  nontopological gray-gray solitons, and bright solitons nestling in the background of nonzero constant amplitude. All these solutions, which are   found in the analytical form, coexist in the system with the same set of model parameters. At the same time, in a properly defined linear limit, the frequency and wavevector of  bright solitons approach the lower-frequency branch of the dispersion relation, while solitons of other types approach the upper polariton branch.   We numerically observe that large-amplitude  bright solitons are prone to oscillatory instabilities which, however, can have rather weak instability increment. Dark-gray and gray-gray solutions are stable in a vast range of parameters. The stability predictions are verified with direct numerical modelling of soliton dynamics.

% The equations can be generalized for vector case to describe propagation of two polarizations in 1D waveguides with TE-TM splitting
%  \begin{subequations}
%	\label{eq:main_vec}
% \begin{eqnarray}
% i(\pz A_{\pm} + \pt A_{\pm}) = -\gamma_{ph} A_{\pm}-\kappa \psi_{\pm}+\sigma A_{\mp}+f_{\pm}(x,t),\\
% i\pt \psi_{\pm} = -\gamma_{ex}\psi_{\pm}-\kappa A_{\pm} + g |\psi_{\pm}|^2\psi_{\pm}+g_x |\psi_{\mp}|^2\psi_{\pm}.
% \end{eqnarray}
% \end{subequations}
%Here indexes $\pm$ stand for clockwise and counter-clockwise polarisations, the TE-TM splitting of the photonic component is accounted by coupling constant $\sigma$. The nonlinear blue shift of spin-up excitons has a contribution proportional to the density of the spin-woen excitons and vice versa. The strength of this effect is defined by the interaction constant $g_x$  To make our theoretical studies close to possible experiments in some part of our investigation we did numerical simulations for the finite losses in the photon and exciton components, these losses are accounted by $\gamma_{ph}$ and $\gamma_{ex}$ correspondingly. Also to study the problem of the excitation of the solitons by externel pulses with finite aperture and duration we include the driving force $f_{\pm}$ in the right hand side of the equation of the photonic component. 

The rest of our paper is organized as follows. In Sec.~\ref{sec:solutions} we present the analytical exact solutions for bright and dark solitons and discuss their spectral stability. In Sec.~\ref{sec:dyn} we present the result of direct numerical modelling of soliton dynamics. Section~\ref{sec:concl} concludes the paper.

\section{Exact moving soliton solutions}
\label{sec:solutions}
\subsection{Construction of solutions}

We are looking for soliton solutions in the moving frame $\z=x-v_st$, where $v_s$ is the velocity (hereafter, the subscript `s' stays for `soliton'). We therefore introduce the following substitutions  $A(x,t) = A_s(\z) e^{-i\delta_s t}$,  and $\psi(x,t) = \psi_s(\z) e^{-i\delta_s t}$. The solitons are characterized by two parameters: velocity $v_s$ and frequency in the moving frame $\delta_s$ (which is generically different from the frequency  $\delta$ in the lab frame). Then   system (\ref{eq:main}) reduces to
\begin{eqnarray}
\label{eq:stat}
i(1-v_s)A_s' + \delta_s A_s = -\kappa \psi_s,\\
\label{eq:stat2}
-iv_s\psi_s' + \delta_s \psi_s = -\kappa A_s + g|\psi_s|^2\psi_s,
\end{eqnarray}
where prime means derivative with respect to  the moving frame coordinate $\z$. Differentiating the second equation of the latter system one more time, one can eliminate the wavefunction $A_s(\z)$ and reduce  the  system to a single equation for $\psi_s(\z)$. 
%The latter  system can be further reduced to a single equation
%\begin{eqnarray}
%-v_s(1-v_s) \psi_s'' + i\delta_s(1-2v_s) \psi_s' + (\kappa^2-\delta_s^2)\psi_s \nonumber\\[2mm] - g\delta_s |\psi_s|^2\psi_s+ ig(1-v_s)(|\psi_s|^2\psi_s)' = %0.
%\end{eqnarray}
%The linear term with the first derivative can be easily  removed with a 
Using a substitution
\begin{equation}
\psi_s(\z) = \phi_s(\z) \exp\left\{-i\z \frac{\delta_s(1-2v_s)}{2v_s(1-v_s)}     \right\},
\end{equation}
where $\phi_s(\z)$ is a new unknown, 
the problem transforms  to
%the  mixed  nonlinear Schr\"odinger equation in the stationary form:
\begin{eqnarray}
\label{eq:phi}
-v_s^2 \phi_s''   + \left(\frac{\kappa^2v_s}{1-v_s} -   \frac{\delta_s^2}{4(1-v_s)^2}\right)\phi_s \nonumber\\[2mm]
+ \frac{g\delta_s }{2 (1-v_s)}|\phi_s|^2\phi_s + igv_s(|\phi_s|^2\phi_s)' = 0.
\end{eqnarray}
%Notice that $\phi_s$ is  a solution of Eq.~(\ref{eq:phi}) with  velocity $v_s>1$, then one can immediately  construct a solution with negative velocity $1-v_s<0$ by replacing  $\delta_s$ with $-\delta_s$ and $\phi_s$ with $\sqrt{-(1-v_s)/v_s}\, \phi_s^*$. 
Next, we use the polar form $\phi_s(\z) = \rho_s(\z) e^{i\Theta_s(\z)}$ and decompose     equation (\ref{eq:phi}) into real and imaginary parts. The latter results in the following relation:
\begin{equation}
\label{eq:Theta}
\Theta_s' \rho_s^2 = \frac{3g}{4v_s} \rho_s^4 + C,
\end{equation}
where $C$ is arbitrary constant of integration. Then the  real part of the polar decomposition becomes
%\begin{eqnarray}
%\label{eq:cq}
%-v_s^2(1-v_s)^2\rho_s'' + %\left(\kappa^2v_s(1-v_s)  %-\frac{\delta_s^2}{4} + %\frac{gCv_s(1-v_s)^2}{2}\right)\rho_s  \nonumber\\[2mm]  - %\frac{g\delta_s(1-v_s)}{2} \rho_s^3
%- \frac{3g^2}{16}(1-v_s)^2\rho_s^5=0.
%\end{eqnarray}
\begin{eqnarray}
\label{eq:cq}
-v_s^2\rho_s'' + \left(\frac{\kappa^2v_s}{1-v_s}  -\frac{\delta_s^2}{4(1-v_s)^2} + \frac{gCv_s}{2}\right)\rho_s  \nonumber\\[2mm]  + \frac{g\delta_s}{2(1-v_s)} \rho_s^3
- \frac{3g^2}{16}\rho_s^5 + \frac{C^2v_s^2}{\rho_s^3}=0.
\end{eqnarray}
In  the particular case $C=0$ the latter equation can be considered as a stationary version of the  cubic-quintic nonlinear Schr\"odinger equation, which is known to support a number of  solutions in the form of bright and   dark solitons,  many of which  can be found in the analytical form, see e.g. \cite{AA,Kivshar,cq}. This observation paves the way towards the systematic construction of analytical  solitons for  the original system (\ref{eq:main}). 
The quintic nonlinearity in (\ref{eq:cq})
is   focusing, while 
%, and its effective strength depends not only on the absolute value of the nonlinear coefficient $g$ but also on the soliton velocity. 
  sign and    effective strength of the  cubic nonlinearity depend both  on the soliton velocity $v_s$ and frequency $\delta_s$.

\begin{figure}%[t]
	\begin{center}
		\includegraphics[width=0.99\columnwidth]{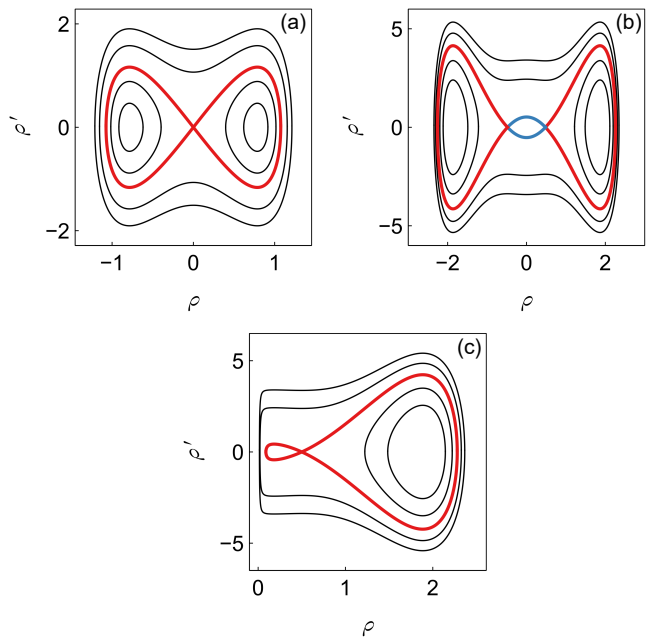}
	\end{center}
	\caption{Phase portraits of Eq.~(\ref{eq:cq}) obtained for $\kappa=g=1$,  $v_s=0.25$, $\delta_s\approx  -0.433$ (a), $\delta_s\approx 1.044$ (b), and $\delta_s=1.07$ (c). Trajectories corresponding to homo- and heteroclinic orbits are highlighted with red and blue colors, respectively. In upper panels $C=0$, and in the lower panel $C\approx 0.051$.}
	\label{fig:portrait}
\end{figure}

Types of existing solutions can be also anticipated from the phase portrait which can be obtained by multiplying Eq.~(\ref{eq:cq}) by $\rho_s'$ and integrating. Representative phase portraits for different combinations of the parameters are presented in Fig.~\ref{fig:portrait}. For $C=0$ and different values of the frequency  $\delta_s $ %\in (-1, 1)$
the system admits   homoclinic orbits which join  the equilibrium $(\rho_s, \rho_s')=(0,0)$ to itself and hence correspond to bright solitons 
[Fig.~\ref{fig:portrait}(a)] or     homoclinic orbits corresponding to bright solitons situated  on a nonzero background coexisting with   and heteroclinic orbits corresponding to dark solitons [Fig.~\ref{fig:portrait}(b)].  For nonzero $C$   [Fig.~\ref{fig:portrait}(c)], the system has   homoclinic orbits of different types that  correspond to  bright solitons  on a nonzero background (with maximal amplitude larger than that of the equilibrium) and to grey solitons (i.e., dips in the uniform background).

If  the amplitude $\rho_s(\zeta)$ is found from Eq.~(\ref{eq:cq}), one can recover the argument $\Theta_s(\z)$ of the corresponding excitonic field using (\ref{eq:Theta}) and then find the photonic component $A_s(\z)$ from Eq.~(\ref{eq:stat2}).  In the particular case $C=0$ for the amplitude of the photonic field   we compute
\begin{eqnarray}
\label{eq:useful}
 |A_s(\z)|^2 = \frac{v_s}{1-v_s} (  \rho_s^2(\z)-\rho_\infty^2) \nonumber\\+ \frac{\rho_\infty^2}{\kappa^2}\left(\frac{g}{4}\rho_\infty^2 - \frac{\delta_s}{2(1-v_s)}\right)^2,
\end{eqnarray}
where constant $\rho_\infty$ is determined by the boundary conditions 
%While complete analytical calculations can be in some cases cumbersome, one can extract some apriori information. In particular, for solutions whose amplitude approaches the same value at both infinities or, more precisely,
\begin{eqnarray}
\label{eq:bc}
\lim_{\z\to \infty}\rho_s^2(\z) &=& \lim_{\z\to -\infty}\rho_s^2(\z) =: \rho_\infty^2,\\[2mm]
\lim_{\z\to \infty}\rho_s'(\z) &=& \lim_{\z\to -\infty}\rho_s'(\z)=0.
\end{eqnarray}
%one can obtain the following useful   relations:
%\begin{eqnarray}
%\label{eq:useful}
% |A_s(\z)|^2 = \frac{v_s}{1-v_s} (  \rho_s^2(\z)-\rho_\infty^2) \nonumber\\+ \frac{\rho_\infty^2}{\kappa^2}\left(\frac{g}{4}\rho_\infty^2 + \frac{\delta_s}{2(1-v_s)}\right)^2,\\[3mm]
% \label{eq:useful2}
% v_s^2[\rho_s'(\z)]^2 = \kappa^2 |A_s(\z)|^2 \nonumber\\ - %\rho_s(\z)^2\left(\frac{g}{4}\rho_s^2(\z) + \frac{\delta_s}{2(1-v_s)}\right)^2.
%\end{eqnarray}
Therefore the squared amplitudes   of $\psi_s(\z)$ and $A_s(\z)$ are proportional, except for an additive constant which is determined by the asymptotic behavior at the infinities.

\subsection{Bright solitons}
\label{sec:bright}
As   is evident from Fig.~\ref{fig:portrait}(a), for $C=0$ in  Eq.~(\ref{eq:cq}) the system supports bright solitons %The best known   class of exact solutions  of cubic-quintic equation (\ref{eq:cq}) corresponds to bright gap solitons 
for which $\rho_\infty=0$ in boundary conditions (\ref{eq:bc}).  As follows from    (\ref{eq:useful}) with  $\rho_\infty=0$,   solutions of this type can only be meaningful   for $v_s\in (0,1)$. Bright soliton  solutions can be written down  in the compact form if ones introduces two auxiliary angles $\alpha \in (0, \pi/2)$ and $\theta \in (-\pi/2, \pi/2)$ and adopts the following parametrization for the soliton frequency and velocity:
\begin{equation}
\label{eq:parametrization1}
v_s = \sin^2\alpha, \quad \delta_s = -\kappa\sin(2\alpha)\sin \theta.
 \end{equation}
Then the following solution can be found  (see also \cite{NatCom15})
\begin{equation}
\label{eq:bright}
\begin{array}{rcl}
\psi_s(\z) & =& \rho_s(\z) e^{ 2i\z\kappa\cot(2\alpha)  \sin \theta   + i \Theta_s(\z) },\\[2mm]
A_s(\z) &=& \tan (\alpha)  \rho_s(\z)  e^{ 2i\z\kappa\cot(2\alpha)  \sin \theta   + i\Theta_s(\z)/3 },
\end{array}
\end{equation}
where
%\begin{eqnarray*}
%\rho^2_s(\z) &=& \frac{4\kappa }{g} \frac{\tan\alpha\cos^2\theta}{\sin \theta + \cosh(4\kappa\cos\theta\csc(2\alpha)\z)},\\[2mm]
%\Theta_s(\z) &=& 
%3\arctan\left(\frac{1-\sin\theta}{\cos\theta} \tanh(2\kappa\cos\theta\csc(2\alpha)\z) \right).
%\label{eq:bright2}
%\end{eqnarray*}
\begin{equation}
\begin{array}{c}
\rho^2_s(\z) = \displaystyle \frac{4\kappa }{g} \frac{\tan\alpha\cos^2\theta}{\sin \theta + \cosh(4\kappa\cos\theta\csc(2\alpha)\z)},\\[6mm]
\Theta_s(\z) = 
\displaystyle 3\arctan\left(\frac{1-\sin\theta}{\cos\theta} \tanh(2\kappa\cos\theta\csc(2\alpha)\z) \right).
\end{array}
\label{eq:bright2}
\end{equation}
Since bright  solitons \rev{(\ref{eq:bright})--(\ref{eq:bright2})} are found in the  frame moving with velocity $v_s$,   the frequency in the lab frame  amounts to $\delta = \delta_s  + 2v_s\kappa \cot(2\alpha)\sin\theta = -\kappa\sin\theta \tan \alpha$, and the spatial wavenumber of soliton tails amounts to $k_s=2\kappa \cot(2\alpha)\sin\theta$.  In the limit $\theta\to \pi/2$ the soliton amplitude tends to zero, and the solution frequency $\delta$ approaches from above the  lower   branch of the dispersion relation, i.e., $\delta = \delta_1(k_s)$, where the polariton dispersion laws $\delta_{1,2}(k)$ are defined in (\ref{eq:dr}). This is shown schematically with the arrow `bs' in Fig.~\ref{fig:dr}. In the limit $\theta\to -\pi/2$, the solution frequency approaches (from below) the  upper   polariton branch $\delta_2(k)$. In this limit, the shape of the solution becomes algebraic:
\begin{equation}
    \lim_{\theta\to-\pi/2} \rho_s^2(\zeta) =   \frac{16\kappa }{g} \frac{\sin(2\alpha)\sin^2\alpha}{\sin^2(2\alpha) + 16\kappa^2\zeta^2}.
\end{equation}

Let us now look into stability of the found bright gap solitons. Using  the standard   linear stability analysis, we consider perturbed stationary solutions in the form
\begin{eqnarray}
\label{eq:pert}
   A(x,t) = e^{-i\delta_s t}[A_s(\z) +  a_1(\z)e^{\lambda t} + a_2^*(\z)e^{\lambda^*t}],\\ 
   \psi(x,t) = e^{-i\delta_s t}[\psi_s(\z) + \ p_1(\z)e^{\lambda t} + p_2^*(\z)e^{\lambda^*t}],
\end{eqnarray}
where $a_{1,2}(\z)$ and $p_{1,2}(\z)$ describe the spatial shapes of the perturbations, and   and complex $\lambda$ characterizes the temporal behavior of the perturbations (positive real part of $\lambda$ means that the perturbations grow and the soliton is therefore unstable). Substituting these expressions in Eq.~(\ref{eq:main}) and keeping only linear (with respect to the small perturbations) terms, we arrive at the following system of linear stability equations which can be treated as an eigenvalue problem for the instability increment $\lambda$:
\begin{equation}
\label{eq:linstab}
\begin{array}{rcl}
i\lambda  a_1  &=& -i(1-v_s)a_1' - \delta_s a_1 - \kappa p_1,\\[3mm]
i\lambda a_2 &=&  -i(1-v_s)a_2'+ \delta_s a_2 + \kappa p_2,\\[3mm]
i\lambda p_1 &=& -\kappa a_1 + iv_s p_1' - (\delta_s - 2g|\psi_s|^2)p_1 + g\psi_s^2 p_2,\\[3mm]
i\lambda p_2 &=& \kappa a_2  - g(\psi_s^*)^2 p_1 + iv_s p_2' + (\delta_s - 2g|\psi_s|^2)p_2.
\end{array}
\end{equation}

\begin{figure}%[t]
	\begin{center}
		\includegraphics[width=0.99\columnwidth]{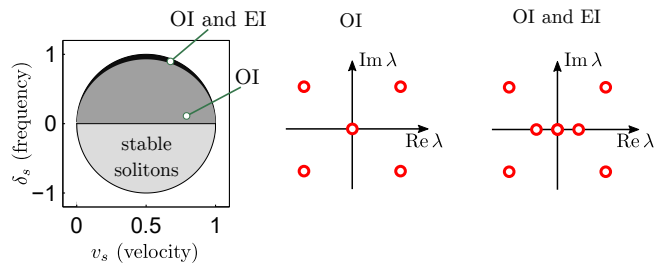}%
	\end{center}
	\caption{\rev{Left panel shows the existence and stability diagram for bright solitons on the plane $(\delta_s, v_s)$ for  $g=\kappa=1$. Light-gray domain corresponds to stable solitons, dark-gray domain corresponds to solitons with oscillatory instabilities (OI), and black sickle-shaped region corresponds to high-frequency solitons  with coexisting   oscillatory and purely exponential instabilities (OI and EI). In white domain bright solitons do not exist.  Two panels on the right show schematically the location of unstable eigenvalues for the two situations. }}
	\label{fig:diadram}
\end{figure}

It is known that dynamics of gap solitons in various setups can be affected by  \emph{oscillatory  instabilities} (OIs) which correspond to   unstable eigenvalues detaching from the edges of the continuous spectrum \cite{Kivshar,Barash,oi}. \rev{By definition \cite{Kivshar}, the instability of this type is   associated with a quartet of complex eigenvalues $(\pm \lambda, \pm \lambda^*)$, see panel `OI' in Fig.~\ref{fig:diadram}.   Another common type of instabilities corresponds to a pair of purely real  eigenvalues $(\lambda, -\lambda)$. The instability of this type can be referred to as the \textit{internal} or \textit{exponential instability} (EI),  see schematic illustration  `OI and EI' in Fig.~\ref{fig:diadram}.}

\begin{figure}%[t]
	\begin{center}
		\includegraphics[width=0.99\columnwidth]{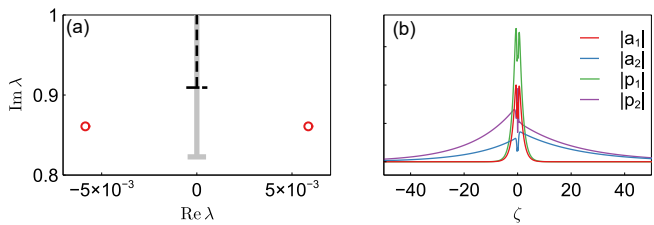}%
	\end{center}
	\caption{(a) A pair of complex eigenvalues $\lambda$ and $-\lambda^*$ (shown with circles) found numerically  in the vicinity of the nearly resonant edges of the continuous spectrum for small negative $\theta$. Two partially overlapping branches of the continuous spectrum are shown with vertical solid gray and dotted black lines, and the edges of the branches are emphasized with   short horizontal bars.  There also exist a pair of eigenvalues  $-\lambda$ and $\lambda^*$ in the lower   half-plane, not shown in the figure. (b) Spatial profile   of the corresponding four-component eigenvector. In this figure, $g=\kappa=1$, $\alpha=\pi/6$, and $\theta=-0.05$. }
	\label{fig:oi}
\end{figure}

\rev{We solve the eigenvalue problem (\ref{eq:linstab}) by approximating the derivatives by finite differences and evaluating the spectrum of the resulting sparse matrix using the MatLab eigs procedure. The fourth-order approximation has been used for the derivatives subject to the zero boundary conditions. Depending on the localization of eigenfunctions, we   have used different computational windows $\z\in[-L, L]$, with $L$ ranging from 20 to 160 and number of grid nodes ranging from $10^4$ to $2\cdot 10^4$. In each case it has been checked that small variations of the grid parameters do not have any essential impact on the outcomes of the computation.} Numerical solution of the linear stability eigenvalue problem (\ref{eq:linstab}) indicates that the oscillatory  instabilities are indeed present in our system.
More specifically, we observe that solitons with $\delta_s<0$ are stable, whereas oscillatory instabilities can be found for $\delta_s>0$. Precise detection of the instability threshold is a numerically challenging problem, because  for   $\delta_s$ close to zero the instability increments are     rather weak, and the decay of the  tails of corresponding unstable eigenmodes  is  extremely slow as $\z$ approaches $\infty$ and $-\infty$.   At the same time, the stability change   at $\delta_s=0$ (which, in terms of parametrization (\ref{eq:parametrization1}) corresponds to $\theta=0$) can be anticipated as one looks at the  continuous spectrum associated with the linear stability system (\ref{eq:linstab}). It has four branches of the continuous spectrum that occupy the following intervals of the imaginary axis:
\begin{eqnarray}
\lambda_{1,2} &\in& i[\kappa \sin(2\alpha)(1 \pm \sin \theta), +\infty),\\[1.5mm]
\lambda_{3,4} &\in&  (-\infty, -\kappa\sin(2\alpha)(1 \pm \sin \theta)]i.
\end{eqnarray}
Exactly at $\theta=0$ the edges of the continuous spectrum coincide pairwise and become resonant, which,  as we conjecture, results in the bifurcation of a quartet of oscillatory instability eigenvalues, with two eigenvalues emerging from   $\lambda =  i \kappa \sin(2\alpha)$ and  two more eigenvalues emerging from  $\lambda =  -i \kappa \sin(2\alpha)$. Weak oscillatory instabilities \rev{(with the instability increment $\RE\, \lambda\lesssim 10^{-2}$)} emerging in the vicinity of the resonant spectrum edges have indeed been observed in our numerical simulations for small positive $\delta_s$ as shown in Fig.~\ref{fig:oi}.  As  $\delta_s$   increases  towards larger positive  values, the  increment    of oscillatory instability grows.

\rev{As pointed out above in this section, apart from the oscillatory instabilities, the system can admit purely exponential instabilities }  associated with a pair of purely real eigenvalues $\pm \lambda \in \mathbb{R}$  in the linearization spectrum. The instability of this type emerges at the moment  when two stable internal modes (i.e., purely imaginary and complex-conjugate  eigenvalues) collide at the origin and then split into a   pair of purely real eigenvalues of opposite sign.  The moment of such an eigenvalue zero crossing can be obtained analytically using the   multiple-scale analysis \cite{Barash,multi}. To this end, we notice that 
%Apart from the Hamiltonian, 
the system (\ref{eq:main}) with zero boundary conditions has two   conserved quantities: $Q = \int_{-\infty}^\infty x (|A|^2 + |\psi|^2)$, and $P = \frac{i}{2} \int_{-\infty}^\infty dx (A \px A^* - A^* \px A + \psi\px \psi^* - \psi^* \px \psi)$.
%\begin{eqnarray}
%Q = \int_{-\infty}^\infty dz (|A|^2 + |\psi|^2),\\
%P = \frac{i}{2} \int_{-\infty}^\infty dz (A \pz A^* - A^* \pz %A + \psi\pz \psi^* - \psi^* \pz \psi).
%\end{eqnarray}
For bright solitons given by the exact solution  (\ref{eq:bright}), these quantities become functions of $\alpha$ and $\theta$:
\begin{eqnarray}
Q_s = \frac{2\tan^2\alpha}{g} (\pi - 2\theta),\\
P_s = \frac{4\kappa \tan\alpha}{g  \cos^2\alpha} [(1+2\cos^2\alpha)\cos \theta - (\pi-2\theta)\sin \theta] .
\end{eqnarray}
The multiple-scale analysis relies on the assumption that if the increment of the newly emerged  exponential instability is small, then the initial stage of the dynamical instability development   results in the adiabatic change of the solitons frequency and velocity, i.e., one can introduce functions $v_s=v_s(T)$ and $\delta_s=\delta_s(T)$, where $T=\epsilon t$ is a `slow time', and $\epsilon$ is a formal small parameter. Respectively, the auxiliary parameters are also to be considered as function of the slow time: $\alpha=\alpha(T)$ and $\theta=\theta(T)$.  Carrying out the corresponding calculations, we observe that  a new small unstable eigenvalue appears in (or disappears from) the linearization spectrum  at the instance when the following condition is satisfied 
%\begin{equation}
%\label{eq:cond1}
% \frac{\partial P_s}{\partial v_s}  \frac{\partial Q_s}{\partial \delta_s} -  \frac{\partial P_s}{\partial \delta_s}  \frac{\partial Q_s}{\partial v_s}=0.
%\end{equation} 
%In the latter equation, the   quantities $P_s$ and $Q_s$ are considered as functions of   free parameters $v$ and $\delta$. However, since the Jacobian of the transformation between $(v_s, \delta_s)$ and $(\alpha, \theta)$ is nonzero:
%\begin{equation}
 %\frac{\partial \alpha}{\partial v_s}  \frac{\partial \theta}{\partial \delta_s} -  \frac{\partial \alpha}{\partial \delta_s}  \frac{\partial \theta}{\partial v_s}\ne 0,
%\end{equation}
%condition (\ref{eq:cond1}) can be replaced with
\begin{equation}
\label{eq:cond2}
\frac{\partial P_s}{\partial \alpha}  \frac{\partial Q_s}{\partial \theta } -  \frac{\partial P_s}{\partial \theta}  \frac{\partial Q_s}{\partial \alpha}=0.
\end{equation} 
Direct computation reduces (\ref{eq:cond2}) to the following simple equation for $\theta$:
\begin{equation}
(\pi^2-4\pi\theta + 4\theta^2-3)\cos\theta + 2(\pi-2\theta)\sin\theta=0.
\end{equation}
Within the interval  $\theta\in (-\pi/2, \pi/2)$, the latter equation has a single root $\theta_\star \approx -1.189$, with $\sin\theta_\star\approx -0.928$. In terms of the soliton frequency and velocity, the found  threshold  corresponds to the following dependence
\begin{equation}
\label{eq:thres}
\delta_{s,\star} = -2\kappa \sqrt{v_s(1-v_s)} \sin\theta_\star = -\kappa \sin(2\alpha)\sin\theta_\star .
\end{equation}

The emergence of a new pair of real eigenvalues as the soliton frequency $\delta_s$ increases above the found threshold value $\delta_{s,\star}$ has been verified by means of direct evaluation of the linear stability eigenvalues. At the same time, since $\delta_{s,\star} $ is well above zero, these new eigenvalues  emerge in the parametric region where  the bright solitons are \emph{already} unstable due to the oscillatory instabilities described above. Therefore, \rev{for a generic initial perturbation}, this additional exponential instability has no significant impact on the overall behavior of the system.

\rev{Results of the stability analysis for bright solitons are summarized in the  diagram shown in Fig.~\ref{fig:diadram}. It shows the domain of existence of bright solitons on the plane $(\delta_s, v_s)$ and demarcates the stability and instability regions. }

%The existence of   found instability threshold has been verified by means of direct evaluation of the linear stability eigenvalues. We have found that bright solitons with the frequencies below the instability threshold [red  area in Fig.~\ref{fig:linstab}(a)] are characterized by a pair of purely real eigenvalues  [Fig.~\ref{fig:linstab}(b)], which implies that the corresponding solutions are unstable. 

%\begin{figure}%[t]
%	\begin{center}
%		\includegraphics[width=0.5\columnwidth]{linstab.png}%
%	\end{center}
%	\caption{(a) Colored  ring-shaped area  on the   $(\delta, v)$-plane corresponds to the domain of existence of bright solitons. Smaller red-colored portion of the ring corresponds to unstable solitons. The boundary between red and blue areas corresponds to the instability threshold in Eq.~(\ref{eq:thres}). (b) and (c) Representative linear stability eigenvalue portraits just above (b) and just below (c) the instability threshold.  Here $\kappa=g=1$. \textcolor{red}{On the vertical axis of the left panel $\delta$ is actually $\delta_s$.} }
%	\label{fig:linstab}
%\end{figure}  

%The obtained stability diagram implies that the algebraically decaying solitons which are situated near the lower dispersion branch (see the limit   $\theta\to-\pi/2$ discussed above in this section) are unstable.

\subsection{Dark-gray solitons and bright solitons on a nonzero    pedestal}
\label{sec:dark}
Phase portrait shown in Fig.~\ref{fig:portrait}(b) indicates that for $C=0$ and sufficiently large frequencies $\delta_s$  the system supports solitons of two more types. First,  the heteroclinic orbit in Fig.~\ref{fig:portrait}(b) correspond  to  dark solitons for which the profile $\rho_s(\z)$ increases monotonically from $-\rho_\infty$ to $+\rho_\infty$ and becomes zero at   some $\z$, where the excitonic wavefunction   $\psi_s(\z)$ has a topological  $\pi$ phase jump. Regarding the corresponding photonic field $A_s(\z)$,  from (\ref{eq:useful}) we observe that for 
%
%As follows from (\ref{eq:useful}), the shape of the $A_0$ in this case  depends essentially on the sign of the  ratio $v/(1-v)$. For 
$v_s\in (0,1)$  the amplitude $|A_s|^2$ corresponds to a dip in the uniform background.  Moreover, the amplitude of the wavefunction nowhere   vanishes, i.e., $|A_s(\z)|>0$ for all $\z$, i.e., corresponds to a nontopological (gray) soliton without the phase jump. 

The second type of  soliton  solutions in the phase portrait Fig.~\ref{fig:portrait}(b) corresponds to bright solitons on nonzero pedestal. For  these solutions both fields $\psi_s(\z)$ and $A_s(\z)$ correspond to   humps on the constant-amplitude background.

%For $v<0$ and $v>1$ the ratio is negative, and the shape of $|A_0|^2$  has a hump at the center of the moving frame $\z=0$.  Therefore, several cases should be considered separately. 
%Regarding the argument of $A_0$,  we can compute
%\begin{equation}
%\arg [\kappa A_0(\z)] = \z \frac{(1-2v)\delta}{2v(1-v)}  + \Theta_0(\z) + \arctan\frac {v\rho_0'(\z)}{\rho_0(\z)\left(\frac{g}{4}\rho_0(\z)^2 + \frac{\delta}{2(1-v)}\right)},
%\end{equation}
%where   $\rho_0'(\z)>0$ can be computed from (\ref{eq:useful2}).

%\subsection{Small-velocity dark-gray solitons}
%\label{sec:smallv}

To present these solutions, for $v_s\in (0,1)$ it is convenient to   introduce the following parametrization
\begin{eqnarray}
\label{eq:param2}
v_s =  \sin^2\alpha, \quad \delta_s = \frac{\kappa}{2}  \sin(2\alpha) (3e^\theta - e^{-\theta}),
\end{eqnarray}
where   $\alpha\in(0,\pi/2)$ and  $\theta>0$,
and the following constants
\begin{eqnarray}
\rho_\infty^2 = \frac{4 \kappa}{g} \tan\alpha \sinh \theta, \qquad  b = \frac{1}{4}(1-e^{-2\theta}), \nonumber \\ p=\kappa \csc(2\alpha)\sqrt{3e^{2\theta}-e^{-2\theta}-2}.
\end{eqnarray}
Then the dark soliton profile reads
\begin{equation}
\label{eq:psidark}
\psi_s(\z) = \rho_s(\z)e^{-i\kappa(3e^{\theta} - e^{-\theta}) \cot(2\alpha)  \zeta+ i\Theta_s(\z)},
\end{equation}
where 
\begin{eqnarray}
\rho_s(\z) = \frac{\rho_\infty \sinh(p\z)}{\sqrt{\cosh^2(p\z) - b}}, \label{eq:dark}\\[3mm]
	\Theta_s(\z) = 6\kappa\sinh\theta\csc(2\alpha) \z \nonumber\\- 3\arctan\left[\sqrt{\frac{1-e^{-2\theta}}{3+e^{-2\theta}}}\tanh(p\z)\right].
	\label{eq:thetadark}
\end{eqnarray} 

For bright solitons on the background,   it is sufficient to redefine
\begin{equation}
    b  = (3+e^{-2\theta})/4,
\end{equation}
and the resulting solution is obtained from (\ref{eq:psidark}) with  
\begin{eqnarray}
\rho_s(\z) = \frac{\rho_\infty \cosh(p\z)}{\sqrt{\cosh^2(p\z) - b}}, \label{eq:dark}\\[3mm]
	\Theta_s(\z) = 6\kappa\sinh\theta\csc(2\alpha) \z \nonumber\\- 3\arctan\left[\sqrt{\frac{1-e^{-2\theta}}{3+e^{-2\theta}}}\coth(p\z)\right]+3\pi/2,
	\label{eq:thetahump}
\end{eqnarray}  
 where (\ref{eq:thetahump}) is only valid for $\z\geq 0$, and for  $\z<0$   the argument must be redefined as an odd function, i.e.,    $\Theta_s(\z):=-\Theta_s(-\z)$.
 
%Wavefunction $A_s(\z)$   can be computed  as follows:
%\begin{enumerate}
%   \item for dark solitons compute  
%    \begin{equation}
%        \rho_s'(\z) = \frac{(1-b)p\rho_\infty\cosh(p\z)}{(\cosh^2(p\z)-b)^{3/2}};
%    \end{equation}
%    and for humps compute
%    \begin{equation}
%        \rho_s'(\z) = -\frac{bp\rho_\infty\sinh(p\z)}{(\cosh^2(p\z)-b)^{3/2}};
%    \end{equation}
%    
%    \item compute 
%    \begin{eqnarray}
%    \psi_s'(\z) = \left[\rho_s' + i\rho_s\left(-\kappa(3e^\theta - e^{-\theta})\cos(2\alpha) + \frac{3g}{4v_s}\rho_s^2\right)\right]\nonumber\\\times e^{-i\kappa(3e^{\theta} - e^{-\theta}) \cot(2\alpha)  \zeta+ i\Theta_s(\z)}
%    \end{eqnarray}
%    
%    \item compute $A_s(\z)$ from  Eq. (\ref{eq:stat2})
%\end{enumerate}
 
The   frequency in the lab frame can be computed from the soliton frequency as $\delta =-\kappa(3e^\theta - e^{-\theta})\cot(2\alpha)v_s+\delta_s = (\kappa/2)(3e^\theta - e^{-\theta})\tan\alpha$. For dark solitons, it is natural to define the small-amplitude limit as $\rho_\infty\to 0$, which in the case at hand corresponds to $\theta \to +0$. In this limit the frequency  $\delta$ approaches the upper polariton branch of the dispersion relation from above as shown schematically with arrow `ds' in Fig.~\ref{fig:dr}.
%(in other words, dark and bright solitons exist in  nonoverlapping  frequency intervals). Nevertheless it is rather remarkably that bright and dark solitons coexist in the system with the same model parameters, i.e., $\kappa$ and $g$. 

Before we proceed to stability of the found solutions, it is important to examine the potential \textit{modulational instability} \rev{\cite{Kivshar}} of the corresponding   constant-amplitude background. Far from the soliton core, the solutions asymptotically transform to constant-ampltidude waves 
  $\psi_0(\z) = \rho_\infty e^{i\Omega \zeta}$,  where 
%\begin{equation}
%\Omega =  \frac{(1-2v_s)\delta_s}{2v_s(1-v_s)} +  \frac{3g}{4v_s} \rho_\infty^2.
%\end{equation}
%For   solutions with $v\in(0,1)$ we obtain
\begin{equation}
\Omega =   \kappa\csc(2\alpha)\left[6\sinh\theta - (3e^\theta-e^{-\theta})\cos(2\alpha)\right].
\end{equation}
For these constant-amplitude solutions, we perform  the standard modulational stability analysis using a substitution similar  to (\ref{eq:pert}) with perturbations $a_{1,2}(\z) = \tilde{a}_{1,2}e^{\pm i \Omega \z + i k\z}$,  $p_{1,2}(\z) = \tilde{p}_{1,2}e^{\pm i \Omega\z  +i  k\z}$,
where $\tilde{a}_{1,2}$ and  $\tilde{p}_{1,2}$  are constants, and real $k$ characterizes the 
%
%It can be shown that the stability of this solution can be described in terms of the characteristic equation
%\begin{equation}
%\det \left(\begin{array}{cccc}
%(1-v) (k+\Omega) + \delta-\Lambda& 0  & -\kappa & 0\\
%0 & (1-v) (k-\Omega) - \delta-\Lambda & 0 & \kappa\\
%-\kappa & 0 & -v (k+\Omega) + \delta + 2g\rho_\infty^2 -\Lambda & g\rho_\infty^2\\
%0 & \kappa &-g\rho_\infty^2 & -v (k-\Omega) - \delta - 2g\rho_\infty^2 -\Lambda
%\end{array}\right)=0,
%\end{equation}
%where real $k$ describes 
the wavenumber of the perturbation. Then the modulational instability eigenvalues $\lambda(k)$ can be found as roots of a quartic characteristic equation. 
%and the imaginary part of  $\Lambda=\Lambda(k)$ is the instability growth rate. 
The exhaustive classification  of all roots   is  a tedious task, but it is possible to describe their behavior  the limit $k\to \infty$. Using computer algebra, one can show, that in this limit  the roots of the characteristic equation have asymptotic behavior as follows
\begin{eqnarray}
\lambda_{1,2}(k) &=& ik\cos^2\alpha  - i c_{1,2} + o(1),\\
\lambda_{3,4}(k) &=& -ik\sin^2\alpha  -i c_{3,4} + o(1),
\end{eqnarray}
where 
\begin{equation}
c_{1,2}=\pm\kappa  e^{-\theta} \cot\alpha, \quad c_{3,4}=\pm \kappa \tan \alpha \sqrt{4-3e^{2\theta}}.
\end{equation}
Therefore the coefficients  $c_{1,2}$ are real, while $c_{3,4}$ are real if and only if $e^\theta < 2/\sqrt{3}$, i.e., $\theta \lesssim 0.144$. If the latter condition is violated, then $\lambda_{3,4}(k)$ acquire a nonzero real part, and  constant-amplitude solutions (and, respectively, the soliton solutions on the constant-amplitude background) become unstable with respect to small-wavelength ($|k|\gg 1$) perturbations. 
%As should be expected, the foundn modulational instability threshold does not depend on the soliton velocity $v_s$ (i.e., on $\alpha$).
Systematic numerical evaluation of  roots of the characteristic equation indicates that below the found  instability threshold  the constant-amplitude waves are stable.

In the region where the modulational instability is absent, we perform an additional search for possible unstable modes by numerical solutions of the linear stability equations (\ref{eq:linstab}).  It indicates that below the modulational instability threshold the dark-gray solitons are stable, except for a narrow parametric interval of weak instabilities centered   around  $v_s=1/2$. Bright solitons on the pedestal are unstable  \rev{even below the modulational instability threshold due to the presence of internal unstable modes associated with localized eigenfunctions of the linear-instability operator}.

\subsection{Gray-gray solitons}
\label{sec:gray}
Phase portrait in Fig.~\ref{fig:portrait}(c) indicates that when the constant of integration $C$ in Eq.~(\ref{eq:cq}) is nonzero, the   bright solitons on the pedestal coexist with  gray solitons, i.e., nontopological dips in the constant-amplitude background, with the amplitude at the dip being nonzero.  While the shape of the bright solitons is similar to those presented in the previous subsection, the gray-gray solitons   constitute a    significant   generalization of the dark-gray solitons presented above. The respective solutions can also be found in analytical form, although the resulting expressions are rather bulky. To simplify the presentation, in this subsection   we set $\kappa=g=1$.  Then the amplitude and phase of the excitonic field is given as  
\begin{eqnarray}
\rho_s(\z) = \rho_\infty \sqrt{\frac{\sinh^2(p\z)+c}{{\cosh^2(p\z) - b}}},
\end{eqnarray}
and
\begin{widetext}
\begin{eqnarray}
 %\\[3mm]
	\Theta_s(\z) = \int_0^\z\left(\frac{3}{4v_s}\rho_s^2 + C\rho_s^{-2}\right)d\z'=\frac{\pi}{2p}\left(\frac{3\rho_\infty^2}{4v_s}\cos^2 s_2 \tan s_1 + \frac{C}{\rho_\infty^2}\cos^2 s_1 \tan s_2\right) \nonumber\\+ \frac{3\rho_\infty^2}{4v_s}\left(\z - \frac{\sin^2s_2\cot s_1}{p}\arctan(\tan(s_1)\tanh(p\z)) - \frac{\cos^2s_2\tan s_1}{p} \arctan(\cot s_1\coth(p\z))\right) \nonumber\\
	+\frac{C}{\rho_\infty^2}\left(\z - \frac{\sin^2 s_1 \cot s_2}{p} \arctan(\tan(s_2)\tanh(p\z)) - \frac{\cos^2 s_1 \tan s_2}{p} \arctan(\cot s_2\coth(p\z))\right),
	\label{eq:generalized}
\end{eqnarray} 
\end{widetext}
where $s_1 = \arcsin \sqrt{b}$ and $s_2 = \arccos \sqrt{c}$.
%If everything is correct, then $\Theta_s(0)=0$. 
For negative $\z$, $\Theta_s(\z)$ must be redefined as an odd function: $\Theta_s(\z) := -\Theta_s(-\z)$. Then the solution can be found as 
\begin{equation}
\psi_s(\z) =  \rho_s(\z) \exp\left\{i\Theta_s(\z) - i\z \frac{\delta_s(1-2v_s)}{2v_s(1-v_s)}     \right\}.
\end{equation}
Analytical expressions for constant $p$, $b$, and $c$ are available in a  computer algebra program, but are too bulky to be presented herein. Instead, in Table~\ref{tab:numbers} we present three numerical sets for gray and bright solitons with different velocities.

%Wavefunction $A_s(\z)$   can be computed  as follows:
%\begin{enumerate}
%    \item   compute  
%    \begin{equation}
%        \rho_s'(\z) = -\frac{\rho_\infty p(b+c-1)\sinh(p\z)\cosh(p\z)}{\sqrt{\cosh^2(p\z)+c-1}(\cosh^2(p\z)-b)^{3/2}}.
%    \end{equation}
%    
%    \item compute 
%    \begin{eqnarray}
%    \psi_s'(\z) = \left[\rho_s' + i\rho_s\left(\frac{\delta_s(1-2v_s)}{2v_s(1-v_s)}  + \frac{3}{4v_s}\rho_s^2 + C \rho_s^{-2}\right)\right]\nonumber\\\times \exp\left\{i\Theta_s(\z) + i\z \frac{\delta_s(1-2v_s)}{2v_s(1-v_s)}     \right\}.
%    \end{eqnarray}
%    
%    \item compute $A_s(\z)$ from  Eq. (\ref{eq:stat2})
%\end{enumerate}

\begin{table*}[]
    \centering
    \begin{tabular}{c|ccccccccc}
         No.&$v_s$ &  $\rho_\infty$  & $\delta_s$ & $C$ & $p$ & $b_{gray}$ & $c_{gray}$ & $b_{bright}$ & $c_{bright}$ \\\hline
         1&$0.25$ & 0.5  & $ 1.07$& 0.0506& $1.0946$ &$0.0466$ & 0.0301 & $0.9534$ & $ 0.9699$\\[1mm]
         2&$0.5$ &  $0.5$  & $ 1.14$ & $0.0557$ & $0.6881$ & $0.0264$ & $0.0899$ & $0.9735$ & $0.9101$\\[1mm]
         3&$0.75$ &  $1$  & $ 1.2$ & $0.7080$ & $1.1396$ & $0.0440$ & $0.2547$ & $0.9560$ & $0.7453$
    \end{tabular}
    \caption{Three numerical sets for gray-gray and bright solitons on constant-amplitude background presented in Sec.~\ref{sec:gray}. Values $v_s$, $\rho_\infty$, $\delta_s$ are exact and values in other columns are approximate; subscripts $gray$ and $bright$ correspond to parameters for the coexisting gray-gray  and bright solitons.}
    \label{tab:numbers}
\end{table*}

%Three sets of parameters for grey solitons (red color) and humps on background (blue color).
%
%\textbf{First set}:
% \begin{eqnarray*}
% \rho_\infty=0.5, \ v_s = 0.25, \ \delta_s = -1.07,\\
% C =   0.050649360483291, \ p=1.094587110827094,\\
% \textcolor{red}{b =   .046640436154287, \ c = .030108139077740},\\
% \textcolor{blue}{b =  .9533595638457123, \ c =   .969891860922260},\\
% \end{eqnarray*}
%
%\vspace{1cm}
%\textbf{Second set}:
%\begin{eqnarray*}
% \rho_\infty=1, \ v_s = 0.75, \ \delta_s = -1.2,\\
% C =   .708033290120886, \ p=1.139600307160625,\\
% \textcolor{red}{b =  .04401232888872617, \ c = .2547146475248614},\\
% \textcolor{blue}{b =  .9559876711112740, \ c = .7452853524751387},\\
% \end{eqnarray*}
% 
% \vspace{1cm}
%\textbf{Third set}:
%\begin{eqnarray*}
% \rho_\infty=0.5, \ v_s = 0.5, \ \delta_s = -1.14,\\
% C = .05567669325356855, \ p=.6881327584511350,\\
% \textcolor{red}{b =  .02646685923953975, \ c =.08986582138990908},\\
% \textcolor{blue}{b =  .9735331407604600, \ c = .9101341786100908},\\
% \end{eqnarray*}

\section{Direct numerical modelling of soliton dynamics}
\label{sec:dyn}

% Now let us consider the development of the instabilities of the soliton solutions discussed above. We start with the bright solitons without a background. Numerical study in Sec.~\ref{sec:bright} has indicated that the solitons are stable  when the soliton frequency $\delta_s$ is positive  [which in terms of the   parametrization  adopted in Eqs.~(\ref{eq:parametrization1}) corresponds to $\theta>0$], but become unstable for negative $\theta$, and the emerging instabilities are initially rather weak, which makes it difficult to detect precisely the instability threshold. These findings agree with the results of numerical simulations of soliton dynamics presented in Fig.~\ref{fig:dr_sim1}. The initial conditions were taken in the form of a soliton perturbed by a weak noise in both $A$ and $\psi$ fields. As one can see for positive $\theta$ the soliton is stable and this complies with the results of the spectral stablity analysis. At negative $\theta$ the soliton becomes unstable and the instability growth rate increases with at higher absolute values of $\theta$, compare the modelling for $\theta=-0.7$ and $\theta=-1$. Relatively weak instabilities (such as those with $\theta=-0.7$) feature a ``radiative'' behavior  which corresponds to a poorly localized in space unstable eigenfunction [exemplified in Fig.~\ref{fig:oi}(b)]. Stronger instabilities, such as those with $\theta=-1$ in Fig.~\ref{fig:dr_sim1},   cause  the  complete destruction of the solitons. 

Now let us consider the dynamical development of the instabilities of the soliton solutions discussed above. \rev{The numerical simulations of the field evolution are done by well known split-step method. At the first step, we solve a linear part of the equation in Fourier space and, at the second step, we solve the nonlinear part. The typical step of the spatial mesh was about $\Delta x =10^{-3}$ and the time step $\Delta t = 10^{-4}$. The boundary conditions are periodic, the width of the simulation window $L$ is much larger than the soliton width ($L\approx 30$). In the case of the solitons on a background, the width of the window was adjusted to provide the continuity of both fields at all points of the simulation interval. It was specially checked that neither spatial no temporal discretizations affected the results of the simulations.  }

\begin{figure}
    \centering
    \includegraphics[width=0.99\columnwidth]{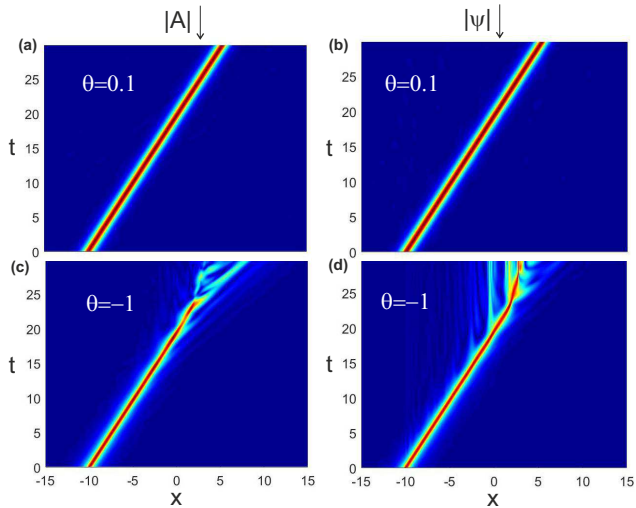}
    \caption{The evolution of a stable (a,b) and an unstable  (c,d) bright solitons. Panels (a,c)  are for $A$ fields and panels (b,d) are for $\psi$ fields.  The initial conditions for the simulations are taken in the form of a soliton solution perturbed by a weak noise. Soliton parameters are $\alpha=0.8$ and $\theta=0.1$ (stable soliton), $\theta=-1$ (unstable soliton).}
    \label{fig:dr_sim1}
\end{figure}

We start with the bright solitons without a background. Numerical study in Sec.~\ref{sec:bright} has indicated that the solitons are stable  when the soliton frequency $\delta_s$ is negative  [which in terms of the   parametrization  adopted in Eqs.~(\ref{eq:parametrization1}) corresponds to $\theta>0$], but become unstable for negative $\theta$, and the emerging instabilities are initially rather weak, which makes it difficult to detect precisely the instability threshold. These findings agree with the results of numerical simulations of soliton dynamics presented in Fig.~\ref{fig:dr_sim1}. The initial conditions were taken in the form of a soliton perturbed by a weak noise in both $A$ and $\psi$ fields. As one can see for positive $\theta=0.1$ the soliton is stable and this complies with the results of the spectral stability analysis. At negative $\theta$ the soliton becomes unstable and the instability growth rate increases with at higher absolute values of $\theta$. Relatively weak instabilities  feature a ``radiative'' behavior  which corresponds to a poorly localized in space unstable eigenfunction. Stronger instabilities, such as those with $\theta=-1$ in Fig.~\ref{fig:dr_sim1}, cause  the  complete destruction of the solitons at relatively short times.

\begin{figure}
    \centering
    \includegraphics[width=0.99\columnwidth]{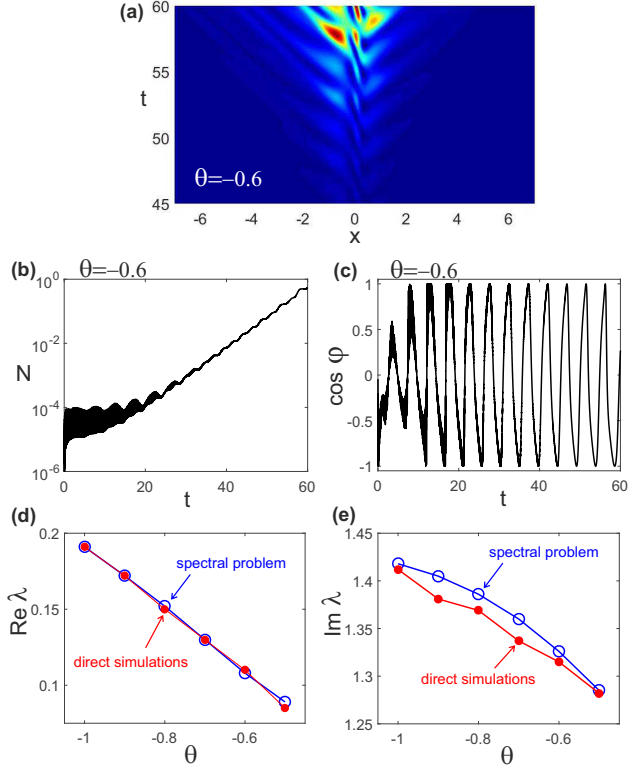}
    \caption{\rev{The evolution of a perturbation growing from the weak noise imposed over an unstable bright soliton solution is shown in panel (a) for $\theta=-0.6$ and $\alpha=0.8$. Panel (b) illustrates the growth of the norm of the perturbation. Panel (c) shows the temporal evolution of the cosine of the mutual phase $\varphi$ of the soliton and the perturbation calculated at the point of the soliton intensity maximum. The values of the real and imaginary parts of the eigenfunctions governing the dynamics of the unstable perturbation are given in panels (d) and (e) for different values of $\theta$. The eigenvalues found by the solution of the spectral eigenvalue problem are shown by open blue circles, the eigenvalues extracted from direct numerical simulations are shown by red circles. The thin lines in these panels are just guides for eye.} }
    \label{add_fig_ref_sgtn}
\end{figure}

\rev{To compare the results of the direct numerical simulations with the predictions of the linear stability analysis obtained from the numerical solution of the  spectral problem (\ref{eq:linstab}), we have extracted the growth rate and the oscillation frequency of the perturbation destroying the solitons. The results of the numerical simulations are summarized in Fig.~\ref{add_fig_ref_sgtn}(a--c) for $\theta=-0.6$. We take the initial conditions in the form of analytical soliton solution perturbed by a weak noise. Then we perform numerical simulation and for each moment of time evaluate  the perturbation on the soliton background. The perturbation is defined as  $\vec U(x,t) =(A(x,t)-A_s(x-\xi)\exp(i\Phi), \psi(x,t)-\psi_s(x-\xi)\exp(i\Phi))^T$ with the phase $\Phi$ and the displacement $\xi$ giving the minimum of the norm $N=\int |\vec U|^2dx$ (here $A_s$, $\psi_s$ are the analytical solution, $A$ and $\psi$ are the fields distributions found by numerical simulations). A typical evolution of the perturbation $\vec U$ is shown in Fig.~\ref{add_fig_ref_sgtn}(a) showing the spatiotemporal evolution of $|\vec U|^2$ in the reference frame moving with the soliton. The oscillatory growth of the perturbation is clearly seen in this figure. Panel (b) of the figure shows the dynamics of the norm $N$ in logarithmic scale. It is seen that after some time, when the growing mode start dominating over other components of the perturbation, the norm $N$ grows exponentially in time. This allows us to extract the growth rate from the results of numerical simulations. We can also calculate the mutual phase between the soliton and the perturbation defined as $\varphi=\arg U_1(x_m,t) $ where $U_1$ is the first component of $\vec U$ and $x_m$ is the coordinate of the maximum of the field intensity distribution.  The temporal behaviour of   $\cos \varphi$ is shown in Fig.~\ref{add_fig_ref_sgtn}(c). At larger $t$ the dependency becomes quasisinusoidal and its inverse period gives an estimate for the imaginary part of the eigenvalue (the frequency) of the growing perturbation.}

\rev{This way we can compare the growth rates (real parts of $\lambda$) and the frequencies (imaginary parts of $\lambda$) of the growing perturbations obtained from direct numerical modeling and from numerical solution of the spectral problem. These values are shown in  Fig.~\ref{add_fig_ref_sgtn}(d),(e) for different values of $\theta$. One can see a good quantitative agreement between the eigenvalues. Therefore  the results of numerical simulations support the conclusion on the presence of oscillatory instabilities in  linear-instability spectra of bright solitons.}

\begin{figure}
    \centering
    \includegraphics[width=0.99\columnwidth]{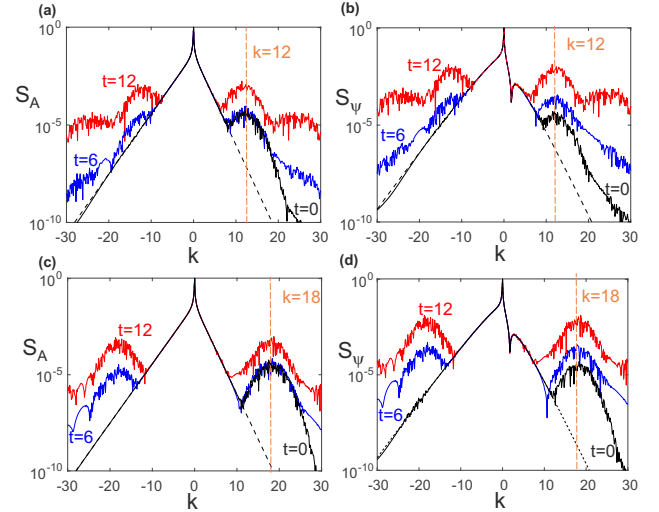}
    \caption{The spectra of $A$ and $\psi$ fields are shown in log scale in panels (a,c) and (b,d) correspondingly. The initial conditions are taken in the form of the dark-gray soliton perturbed by a weak noise. The spectrum of the noise is a Gaussian function of the width $w_s=3$ centered at $k=12$ in (a,b)  and $k=18$  in (c,d). The positions of the centers of noise spectra are marked by the dashed orange lines labeled as $k=12$ and $k=18$. The dark line shows the spectra of the initial fields, the black dashed curves  are the soliton spectra with the parameters $\alpha=\pi/6$ and $\theta=0.25$. The spectra at $t=6$ and $t=12$ are shown by the  blue and the red lines. }
    \label{fig:dr_sim3}
\end{figure}

\begin{figure}
    \centering
    \includegraphics[width=0.99\columnwidth]{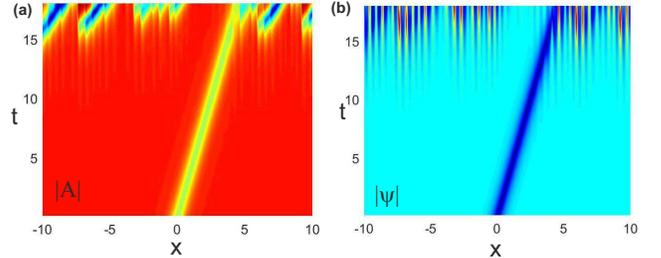}
    \caption{The temporal evolutions of $A$  and $\psi$ fields are shown in panels (a) and (b). The initial distributions of the fields are taken in the form of the soliton solution with $\alpha=\pi/6$ and $\theta=0.25$ perturbed by a weak noise having the Gaussian spectrum of width $w_s=3$ centered at $k=12$. }
    \label{fig:dr_sim4}
\end{figure}

Now we proceed to various solitons that nestle on a background of constant  nonzero amplitude. According to the results of Sec.~\ref{sec:dark}, eventual instability of these solutions can originate either in the modulational instability of the constant-amplitude background or in   internal unstable modes of the soliton itself. In order to illustrate the instability of the former type, we consider %  Now let us consider the stability of dark solitons. We start with the case of 
a dark-gray soliton  nestling on an unstable background. In this case the unstable modes can be characterized by a wavevector because far away of the soliton core the asymptotical behavior of the eigenfunctions corresponds to plane waves. The spectral analysis shows that all modes with relatively high $k$ are  unstable. So to demonstrate the instability and be sure that the numerical method is valid, we take the noise with localized spatial spectrum and check that  at large   simulation times we do not see the growth of the modes with very high $k$ and thus our discretization does not affect the results of the simulations.  To this end, we prepare the initial random perturbation   as follows: we take random field distribution, calculate its Fouirer transform, multiplied it with a Gaussian function centered at some $k$, and then calculate  the inverse Fourier transform. The spectra of $A$ and $\psi$ fields at different propagation times are shown in Fig.~\ref{fig:dr_sim3}. The spectra of the initial distribution are shown by black lines. The spectra of pure soliton solutions are shown as a reference by a dashed black lines. We observe that  the perturbation grows and the satellite spectral lines appear. At large times the instability destroys the background and, correspondingly, destroys the solitons, see   Fig.~\ref{fig:dr_sim4} illustrating this process.

\begin{figure}
    \centering
    \includegraphics[width=0.99\columnwidth]{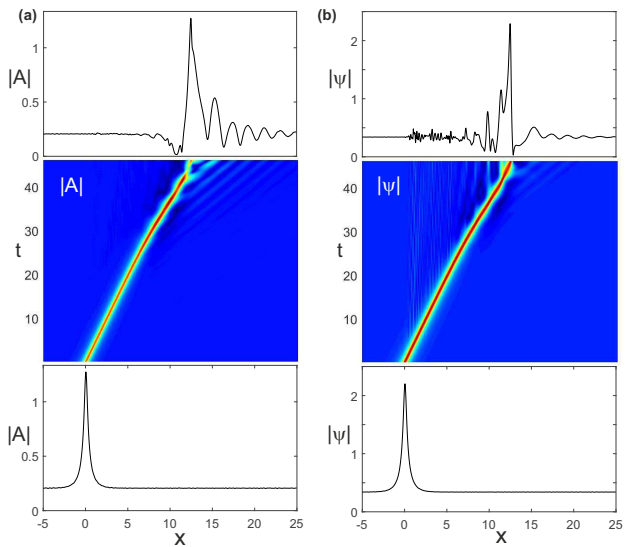}
    \caption{The decay of a bright soliton on the background. Panel (a) shows the initial distribution of $A$ field (lower part), the evolution of the field (in the middle), and the field distribution at $t=46$ (the upper part). Panel (b) shows the same for $\psi$ field.  The initial conditions for the simulations are taken in the form of a soliton solution perturbed by a weak noise. Soliton parameters are $\alpha=\pi/6$ and $\theta=0.05$ [see Eqs.~(\ref{eq:param2}) for the adopted parameterization].}
    \label{fig:dr_sim2}
\end{figure}

We have also examined the stability of  bright solitons on a pedestal that are another kind of possible localized solutions.\rev{ These solitons are unstable even below the modulational instability threshold because of the presence of spatially localized unstable    modes}.  This dynamical instability  is illustrated in   Fig.~\ref{fig:dr_sim2} showing the decay of these solitons into dispersive waves envelope overlapped with the background.

\begin{figure}
    \centering
    \includegraphics[width=0.99\columnwidth]{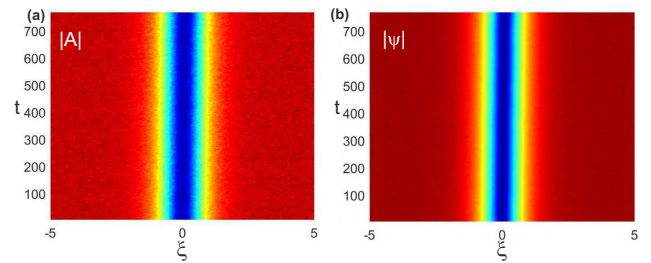}
    \caption{The evaluations of $A$ (a) and $\psi$ (b) fields are shown for a generalized grey soliton. The horizontal axis $\xi=x-v_st$ where $v_s$ is the velocity of the analytically found soliton. The numerical  soliton 
    parameters correspond to No.~1 in Table.~\ref{tab:numbers}.
    %are $\rho_{\}infty} = 0.5$, $v_s = 0.25$, $\delta_s = 1.07$, $C=0.050649360483291$, $p=1.094587110827094$, $b=0.046640436154287$, $c=0.030108139077740$. 
    }
    \label{fig:grey_G_S_modelling}
\end{figure}

\begin{figure}
\centering
\includegraphics[width=0.99\columnwidth]{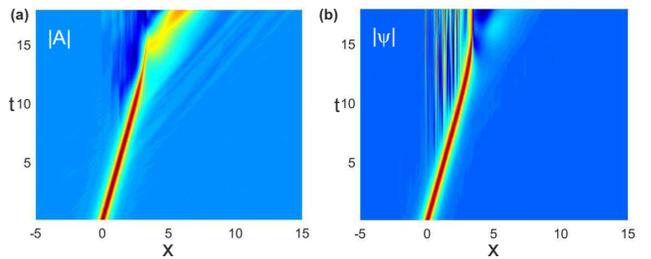}
\caption{The evaluations of $A$ (a) and $\psi$ (b) fields are shown for a 
generalized bright soliton.  The numerical  soliton 
    parameters correspond to No.~1 in Table.~\ref{tab:numbers}.
    %The parameters of the soliton are 
%$\rho_{\infty} = 0.5$, $v_s = 0.25$, $\delta_s = 1.07$, $C=0.050649360483291$, $p=1.094587110827094$, 
%$b=0.9533595638457123$ and $c=0.969891860922260$.
}
\label{fig:bright_G_S_modelling}
\end{figure}

To continue, we have investigated the behaviour of generalized, gray-gray solitons found in Sec.~\ref{sec:gray}. The numerical simulations confirm that these solitons  can be stable.  Fig.~\ref{fig:grey_G_S_modelling} shows the evolution of a perturbed grey-grey soliton in a reference frame moving at the velocity of the unperturbed solitons. The use of the moving reference frame allows to notice small variations in the shape and small displacement of the solitons. However the perturbations do not grow indicating that the solitons are stable. This means that the results of direct numerical simulations are in agreement with the prediction of the linear spectral analysis discussed above. Thus one can conclude that the gray-gray solitons are stable provided that their constant-amplitude background is stable.  Finally, we have also checked   the dynamics of the generalized bright solitons for the three   cases listed in Table~\ref{tab:numbers}. The simulations show that all bright solitons are unstable and   get destroyed in a similar manner, see Fig.~\ref{fig:bright_G_S_modelling} showing a typical spatial-temporal evolution of the $A$ and $\psi$ fields. The instability is presumably of radiative kind, at first the preturbations growth in $\psi$ field, cascading and making the spatial spectrum wider. Then it affects the $A$ field causing complete destruction of the soliton.

% Finally, in  Fig.~\ref{fig:grey_G_S_modelling} we illustrate the possibility of stable evolution of generalized, gray-gray solitons found in Sec.~\ref{sec:gray}.
%In the same time the grey solitons nestling on the same backgrounds look stable, see It is true that the numerical modeling cannot be an ultimate prove of stability but 
% Numerical simulations reveal that when the constant-amplitude  background is stable,  the gray solitons propagate without any noticeable changes for quite long times.
%\begin{figure}
%    \centering
%    \includegraphics[width=0.99\columnwidth]{fig_GreyG.eps}
%    \caption{The evaluations of $A$ (left column) and $\psi$ (right column) fields are shown for different 
% generalized grey solitons. The horizontal axis $\xi=x-v_st$ where $v_s$ is the velocity of the analytically found % % soliton. The soliton parameters are $\rho_{\}infty} = 0.5$, 
% $v_s = 0.25$, $\delta_s =-1.07$, $C=0.050649360483291$, $p=1.094587110827094$, $b=0.046640436154287$, 
% $c=0.030108139077740$ for the upper row (panels (a) and (b) ); 
% $\rho_{\infty} = 1$, $v_s = 0.75$, $\delta_s =-1.2$, $C=0.708033290120886$ $p=1.139600307160625$, 
% $b=0.04401232888872617$, $c=0.2547146475248614$ for the middle row (panels (c) and (d))
%and 
% $ \rho_{\infty} = 0.5$, $v_s = 0.5$, $\delta_s =-1.14$, $C=0.05567669325356855$, $p=0.6881327584511350$, $b=0.02646685923953975$, $c=0.08986582138990908$ for the lower row (panels (e) and (f)). }
%    \label{fig:grey_G_S_modelling}
% \end{figure}

% \begin{figure}
%    \centering
%    \includegraphics[width=0.99\columnwidth]{fig_brightG.eps}
%    \caption{The evaluations of $A$ (left column) and $\psi$ (right column) fields are shown for different 
% generalized bright solitons. The parameters of the solitons are 
%    $\rho_{\infty} = 0.5$, $v_s = 0.25$, $\delta_s =-1.07$, $C=0.050649360483291$, $p=1.094587110827094$, 
% $b=0.9533595638457123$ and $c=0.969891860922260$ for the upper row (panels (a),(b));
%    $\rho_{\infty} = 1$, $v_s = 0.75$, $\delta_s =-1.2$, $C=0.708033290120886$, $p=1.139600307160625$, 
% $b=0.955987671111274$ and $c=0.7452853524751387$ for the middle row (panels (c),(d));
%    $\rho_{\infty} = 0.5$, $v_s = 0.5$, $\delta_s =-1.14$, $C=0.05567669325356855$, $p=0.6881327584511350$, 
% $b=0.97353314076046$ and $c=0.9101341786100908$ for the lower row (panels (e),(f)). }
%    \label{fig:bright_G_S_modelling}
% \end{figure}

\section{Conclusion}
\label{sec:concl}

In this work, we have thoroughly investigated, analytically and numerically, different families of the solitons that exist in the experimantally relevant model describing the evolution of optical pulses in the conservative systems with strong light-matter coupling. In particular, we found analytically the solutions for bright solitons on zero and non-zero backgrounds (solitons on pedestal), Ising-like dark solitons having a point where the excitonic field intensity is exactly zero and the phase of the field shifts by $\pi$ and grey (Bloch-like) solitons where the intensity has a deep and the phase of the field is continuous and rotates in the soliton core.  The corresponding two-component solutions can be termed to as dark-gray (`half-topological') and gray-gray (nontopological) solitons.  All found solutions coexist in the system but, in a properly defined linear limit, detach from different polariton branches of the dispersion law: bright solitons bifurate from the lower branch towards the gap, and all other solutions detach from the upper branch.

We have found that stability of bright solitons on zero background can be affected by oscillatory (radiative) instabilities which emerge when the soliton frequency becomes positive. The oscillatory instability increment is initially weak, but grows distinctively as the soliton frequency increases. Instability of the solitons on the nonzero background  can develop either from the modulational instability of the constant-amplitude waves or from the internal unstable modes. All examined  bright-bright solitons are unstable, while dark-gray and gray-gray solitons are stable in vast parametric regions below the modulational instability threshold. The stability predictions, including the prominent role of the radiative instability, have been verified in direct numerical modelling of soliton dynamics.

%The stability analysis reveals that the solitons can be destabilized by two different instabilities. The first one occurs when two discrete eigenvalues in the spectrum gap collide producing an unstable mode with localization length comparable with the size of the soliton. The instability of the second kind appears when two discrete eigenvalues collide with the edge of the continuum producing a quartet of eigenvalues with non-zero real parts. The eigenvalues with positive real parts are responsible for the radiative instability. In this case the unstable mode is poorly localized in the vicinity of the instability threshold, this makes it difficult to find the instability threshold precisely. 

%We found out that the radiative instability dominates in the case of the considered gap-solitons. By numercial simulations we demonstrated that bright solitons of low amplitudes are stable but at some point pet destabilized by radiative instability destroying the solitons. At even higher intensity another instability appears but it has an increment lower than that of radiative instability. So the radiative instability always dominate. Bright solitons on the background are always unstable.
%Dark and grey solitons happen to b unconditionally stable probided that they nestle on a stable background. 

The results of the paper shed light on possible localized solutions
%, i.e., gap solitons, 
that may exist in the conservative system with strong light-matter coupling when only material excitations are nonlinear. The variety of found analytical solutions  can be used as a starting point for developing a perturbation theory for the dissipative and driven-dissipative systems where the dissipative terms can be considered as perturbations. We believe that this will facilitate the theoretical studies of the hybrid systems with light-matter interactions.

\begin{acknowledgments}
 The research  was supported by   Priority 2030 Federal Academic Leadership Program. 
\end{acknowledgments}

\end{document}